\documentclass{aastex}          
\usepackage{spr-astr-addons}    
\usepackage{natbib}             
\usepackage{multirow}
\usepackage{bigstrut}

\newcommand{\rmn}{\textrm}
\begin{document}

\title{
A technique for estimation of starburst masses and ages in luminous compact 
galaxies
}

\shortauthors{S. L. Parnovsky, I. Y. Izotova}

\shorttitle{
Estimating of starburst masses and ages in luminous compact galaxies
}

\author{S. L. Parnovsky}
\affil{Astronomical Observatory of Taras Shevchenko Kyiv National University\\
Observatorna str., 3, 04053, Kyiv, Ukraine\\
tel: +380444860021, fax: +380444862191\\ e-mail:par@observ.univ.kiev.ua}
\email {par@observ.univ.kiev.ua}

\author{I. Y. Izotova}
\affil{Astronomical Observatory of Taras Shevchenko Kyiv National University\\
Observatorna str., 3, 04053, Kyiv, Ukraine\\
tel: +380444860021, fax: +380444862191\\ e-mail:izotova@observ.univ.kiev.ua}
\email {izotova@observ.univ.kiev.ua}

\begin{abstract}
We propose a technique for estimation of the mass $m$ of the young 
stellar population and the starburst age $T$ in luminous compact galaxies 
(LCGs). For this purpose we use LCG H$\alpha$ emission line 
luminosities from the Sloan Digital Sky Survey (SDSS) spectra and {\sl Galaxy
Evolution Explorer} ({\sl GALEX}) FUV and NUV continuum luminosities.
The method is
intended for quick estimation of $m$ and $T$ in large galaxy
samples and does not require spectral energy distribution (SED) fitting.
Estimated $m$ and $T$ for the sample of about 550 LCGs
are compared with the same values derived from the SED fitting
in the wavelength range $\lambda$$\lambda$ 3800 -- 9200\AA. 
We obtain the average 
differences in $\log m$ and $T$ of 0.27 and 0.87 Myr, respectively.
This technique could
be used for selection of galaxies with desired ranges
of $m$ and $T$ or for reducing a range of
parameter variations in SED fitting.
\end{abstract}


\section{Introduction}\label{s:Introduction}

The goal of this article is to describe a method for estimating 
some starburst parameters of luminous compact galaxies (LCGs). 
LCGs are characterised by a strong burst of star formation \citep{I11} and
have the properties similar to those in so called ``green pea'' (GP) galaxies
discussed by \citet{C09}. 

GPs were selected from the Sloan Digital 
Sky Survey (SDSS) images \citep{A09} in the framework of the Galaxy Zoo project as compact objects
with a green color. This color indicates the presence of strong 
[O {\sc iii}] $\lambda$5007 emission line redshifted to the SDSS $r$ band
in galaxies with redshifts $z$ $\sim$ 0.1 -- 0.3.

On the other hand, LCGs were selected from the SDSS 
spectra as the objects with strong emission lines and they are characterised by
a wider range of redshifts $z$ $\sim$ 0.02 -- 0.6 \citep{I11}. 
Therefore, depending on
the redshift, LCGs on composite SDSS images can be blue, pink, white, green,
brown while other characteristics are the same as those in GPs. Selection 
criteria and derived global characteristics of LCGs are described in 
\citet{I11}. Briefly, these criteria were as 
follows: high equivalent width and large luminosity of the H$\beta$ emission 
line (EW(H$\beta$) $\geq $ 50\AA, $L$(H$\beta$) $\geq $ 
3$\times$10$^{40}$ erg s$^{-1}$, respectively); well-detected 
[O {\sc iii}] $\lambda $4363 {\AA} emission line in galaxy spectra, with a 
flux error less than 50\%. Only star-forming galaxies with typical angular 
diameters $\le 10\arcsec$ were selected. All these criteria only select
objects with strong emission lines in their spectra and thus with 
young starburst ages 3~-~5 Myr for which the accurate abundance determination 
using the direct method is possible. 
\citet{I11} concluded that GPs sample is a subset of a larger sample of 
$\sim$ 800 LCGs in a relatively narrow range of redshifts.

General properties of GPs and LCGs seem to be uncommon in the nearby universe 
suggesting a short and extreme phase of their evolution \citep{A10} and may 
represent the star formation mode prevailing in the early Universe \citep{C09}.
LCGs and GPs are characterised by low oxygen abundance (of $\sim$ 20 \% solar)
\citep{A10,I11,P12,H12}, low stellar mass of 
$\sim 10^{8.5}$ -- $10^{10}M_\odot$, 
high star formation rates (SFR $\sim 10~M_\odot$yr$^{-1}$) 
and high specific star formation rate 
(up to $\sim 10^{-8}$yr$^{-1}$) \citep{C09,I11}. These features 
place LCGs and GPs between nearby blue compact dwarf (BCD) galaxies on one 
side and high-redshift UV-luminous Lyman-break galaxies (LBG) and Ly$\alpha$ 
emitters on the other side.

\citet{PII} used the regression analysis to study the dependence of the 
LCG monochromatic luminosities in the 
{\sl Galaxy Evolution Explorer (GALEX)} far-UV 
(FUV, $\lambda_{\rm eff}$=~1528\AA) and near-UV 
(NUV,  $\lambda_{\rm eff}$ =~2271\AA) bands and
the H$\alpha$ emission line luminosities on parameters derived 
by \citet{I11} for LCGs. \citet{I11} used 
all LCG spectra and Monte Carlo simulations to fit
spectral energy distributions (SED)
in the wavelength range $\lambda\lambda$3800 -- 9200\AA. As far as the spectral contribution 
of the gaseous
continuum in LCGs is high, it was fitted first using
equivalent width EW(H$\beta$) of the H$\beta$ emission line and subtracted
from the observed spectrum prior to the fitting of the stellar continuum. 
It is assumed that young stars were formed in a single burst
with the age which is varied in the range $0 - 10$ Myr. It is also adopted that
old stars were formed continuously with a
constant star formation rate. The fitting provided the set of parameters
for each galaxy, namely the masses of
young $m$ and old $M$(old) stellar populations,
the age of a starburst $T$, and the lower
($t_2$) and upper ($t_1$) limits for the age of old stars 
\citep[see more details in ][]{I11}.
\citet{PII} found that only three parameters, namely oxygen abundance 12+logO/H,
the mass $m$ and the age $T$ of the young stellar population have the 
statistically significant impact on the luminosity.
More specifically, LCG luminosities $L$(FUV), $L$(NUV) and
$L$(H$\alpha$) are proportional to the mass $m$ while the ratio $L/m$ depends 
on the age of the starburst $T$. Thus these parameters are fundamental 
to study the evolution of LCG luminosity.

In the present paper we propose a simple technique for estimating 
of the young stellar population mass $m$ and age $T$ from the LCGs 
luminosities in three wavelength bands without invoking SED fitting.
We give a brief description of the method in
Section \ref{s:Idea} and provide details in Section \ref{s:Three}. 
In Section \ref{s:Alternative} we describe one of the variants of the 
proposed method. 
In Section \ref{s:Sample} we describe the sample and its subsamples
considered in this paper. 
The parameters,
which are necessary for $m$ and $T$ estimation,
are discussed in 
Section \ref{s:Parameters}. In Section \ref{s:Precision} we compare the
values $m$ and $T$ derived with technique proposed in this paper to those 
obtained by \citet{I11} from fitting
the SED of galaxies, and estimate the accuracy of the proposed method.
In Section
\ref{s:Without} we consider the accuracy of the method excluding the oxygen 
abundances [O] from the parameter set. 
The accuracy of the $m$ and $T$ estimation from luminosities 
only at two wavelengths is discussed in Section \ref{s:Two}.
In Section \ref{s:Summary} we summarise the main results of this paper.

For distance estimates we assume $H_0$ = 75 km s$^{-1}$ Mpc$^{-1}$.

\section{Description of the method}\label{s:Idea}

We describe the method using the results of the paper by \citet{PII}.
To simplify the 
way of comparing $L($H$\alpha)$, $L$(FUV) and $L$(NUV), \citet{PII}
used the calibration for SFRs averaged over a reasonable time
scale for different SFR tracers and defined by \citet{K98} as
\begin{equation}\label{e1}
\rmn{SFR}=k \times L.
\end{equation}
The conversion factors $k$ between the SFR and 
$L$(H$\alpha$), $L$(FUV) and $L$(NUV) as well as other star formation tracers 
are discussed by \citet{Ca12}. \citet{K98} derived the coefficient 
$k$ of $7.9 \times 10^{-42}$ for the H$\alpha$ luminosity and
$1.4 \times 10^{-28}$ for the FUV and NUV
luminosities assuming the solar 
metallicity, the initial mass function with the power-law index  2.35 and 
mass limits of 0.1 and 100 $M_\odot$ \citep{S55}, and 
expressing SFR in $M_{\odot}$ yr$^{-1}$,
$L($H$\alpha)$ in erg s$^{-1}$, $L$(FUV) and 
$L$(NUV) in erg s$^{-1}$ Hz$^{-1}$.
We use these coefficients to calculate 
SFR values 
for the sample of LCGs.

According to \citet{PII}
the statistical dependence of the luminosity of an individual galaxy on 
its parameters has the form 
\begin{equation}\label{e2}
SFR= m\left(Af(t)+B\Delta\right)
\end{equation}
with
\begin{equation}\label{e3}
f(t)=\left\{
\begin{array}{rl}
1&\text{if $t<0$;}\\
\exp(-pt)&\text{if $t>0$,}\\
\end{array}\right.
\end{equation}
where $t=T-T_0$, $\Delta$= [O] -- [O]$_0$, 
[O]$\equiv$ 12 + log(O/H) is the oxygen abundance, and $A, B$ and $p$ are
constants. We also use two parameters $T_0=3.2$ Myr and
[O]$_0=8.1$. \citet{PII} chosen the parameter [O]$_0$ as the mean oxygen 
abundance either of the entire sample or its subsamples. In the present paper
few subsamples are considered, therefore we fix the value 
[O]$_0$=8.1 which is the mean oxygen abundance of the entire sample. 
According to Eq. \ref{e3}, the SFR and the luminosity of the galaxy 
are constant for $T<T_0$ and decrease exponentially for $T>T_0$.
This dependence on time is applicable to H$\alpha$, FUV and NUV luminosities
with the same value $T_0=3.2$ Myr corresponding to the lifetime of 
the most massive stars with a mass $\sim$ 100 $M_\odot$. 
The similar dependence in Eq. \ref{e2} and the same value of $T_0$ for strongly
star-forming LCGs revealed by \citet{PII} imply that H$\alpha$, FUV and NUV 
emissions in LCGs are produced by the same young stellar populations.
These facts justify the applicability of the $m$ and $T$ estimation
from the luminosities in different bands.

The ionising radiation responsible for H$\alpha$ emission is produced by the 
most massive O stars, while B stars with longer lifetimes 
contribute to the FUV and NUV bands, in addition to O stars. 
Therefore, the FUV and NUV luminosities decrease with time more slowly 
compared to the H$\alpha$ luminosity, satisfying inequalities 
$p(H\alpha)>p(FUV)>p(NUV)$.

Combining equations (\ref{e1}) and (\ref{e2}) we obtain relations
\begin{equation}\label{e3a}
L= m\left(A^*f(t)+B^*\Delta\right), A^*=A \times k, B^*=B \times k.
\end{equation}
Thus, only two parameters $A^*$ and $B^*$ for the determination 
of $m$ and $T$ are needed.
We note that it is not necessary to know the correct values of $k$ for 
H$\alpha$ line and UV radiation. Overrating the real value $k_{real}$ by a 
factor $w$ and using the value $k=w\times k_{real}$, we obtain 
$A=w^{-1}\times A_{real}$ and $B=w^{-1}\times B_{real}$ instead of correct 
values $A_{real}$ and $B_{real}$. However, 
$A^*$ and $B^*$ remain unchanged and overrating does not affect the 
$m$ and $T$ values.

For comparison of LCG SFRs with SFRs derived in different studies for galaxies
of different types we adopt the coefficient $k$ in Eq. \ref{e1} proposed 
by \citet{K98} for
continuous or quasi-continuous star formation which is common in big galaxies 
with frequent starbursts. This value of $k$ may not be used for LCGs where star
formation occurs in strong and rare bursts. However, the variations of $k$ do 
not influence the $m$ and $T$ values, as discussed above. 

The dependence in Eq. \ref{e2} for emission in the H$\alpha$ line, FUV or NUV 
continuum is characterised by only three parameters $A$, $B$ and $p$. 
We introduce the index $i=1,2,3$ denoting the H$\alpha$, FUV and NUV spectral
bands, respectively. Luminosity decay times for H$\alpha$ line and in the 
FUV or NUV ranges are given
by $\tau_i=p_i^{-1}$. The relations $p_1>p_2>p_3$ and respectively 
$\tau _1<\tau _2<\tau _3$ are satisfied for our LCG subsamples 
(see Table \ref{t1}). 
The ratios of $L$(H$\alpha$) to $L$(FUV) or $L$(NUV) and $L$(FUV) to $L$(NUV) 
decrease exponentially at $T>T_0$. Therefore, one can determine the value of 
$T$ using any of these ratios.

The details of this procedure are described in Section~\ref{s:Two}. To apply
the proposed method one has to know the set of parameters $A_i, B_i$ and 
$p_i$ for the FUV or NUV bands and/or the H$\alpha$ line. 
In Section \ref{s:Parameters} we describe how to obtain these parameters.
Each set of parameters for the H$\alpha$ line and the FUV and NUV range is the 
common one for the entire LCG sample. To estimate $m$ and $T$ for 
individual galaxy one has also to know its luminosities in the FUV or NUV
bands and/or in the H$\alpha$ line, as well as its oxygen abundance [O].

One cannot determine the starburst age $T$ if $T<T_0$, where
luminosities and their ratios do not depend on the time.
However, we can identify such a situation and this information is sufficient 
for the determination of the initial luminosities $L_0$ at the starburst age
$T = 0$ introduced by \citet{PII}. 
The initial H$\alpha$,
FUV and NUV luminosities can be obtained from current luminosities
and the value $f(T)$ from Eq. \ref{e2}. 
If we know the value of $T>T_0$ or
the fact that $T<T_0$, the value of $m$ can be derived from the
H$\alpha$, FUV and NUV luminosities using Eq. \ref{e2}.

The estimation of $m$ and $T$ from the luminosities only in two wavelength 
bands is not very accurate due to the influence of measurement errors and 
deviations from the statistical relation in Eq. \ref{e2}. On the other hand,
the estimation of $m$ and $T$ values from the luminosities in all three 
wavelength bands is more accurate. It is described in Section \ref{s:Three}.

\section{The estimation of $m$ and $T$ values from the luminosities at three
wavelengths}\label{s:Three}

Using Eq. \ref{e2} one can 
determine $m$ and $T$ for each galaxy minimizing the expression
\begin{equation}\label{e4}
U=\sum_{i=1}^3\left(\frac{SFR_i-m(A_if_i(t)+B_i\Delta)}{\sigma_i}\right)^2.
\end{equation}
where $U$ is the sum of the weighted deviations of H$\alpha$, FUV and NUV
luminosities.
Here we use different statistical weights for data at different wavelengths.
They are chosen inversely proportional to the variances $\sigma_i^2$, were 
$\sigma_i$ are mean dispersion for the approximation in Eq. \ref{e2} at each of
three wavelengths. These values are also obtained in Section 
\ref{s:Parameters}.

The derivative of $U$ with respect to $m$ should be equal to zero. This 
requirement yields the condition
\begin{equation}\label{e5}
\displaystyle m=\frac{\sum_{i=1}^3SFR_i(A_if_i+B_i\Delta)\sigma_i^{-2}}
{\sum_{i=1}^3(A_if_i+B_i\Delta)^2\sigma_i^{-2}}.
\end{equation}
If $t=T-T_0>0$, we have $df_i(t)/dT=-f_ip_i$ from Eq.~\ref{e3} and the 
condition $\partial U/\partial T=0$ has the form
\begin{equation}\label{e6}
\sum_{i=1}^3SFR_iA_if_ip_i\sigma_i^{-2}=m \sum_{i=1}^3(A_if_i+B_i\Delta)A_if_ip_i\sigma_i^{-2}.
\end{equation}
If $t>0$, the numerical solution of Eq. \ref{e6} together with
Eq. \ref{e5} and $f_i=\exp (-p_it)$ provides an estimated value of $t$.
Then we derive $T_e=T_0+t$ while Eq. \ref{e5}
provides an estimation of $m$. In the case of $t<0$ ($T_e\le T_0$) we
derive the estimation of $m$ from Eq. \ref{e5} adopting $f_i=1$. In
such a way we obtain the estimations of the age $T$ of the starburst, which 
we denote $T_e$, and the mass $m$ of the young stellar population, which we 
denote $m_e$ (subscript $e$ means estimated).

\section{Alternative version of the method of $m$ and $T$ determination}\label{s:Alternative}

Consider an alternative version of the method which slightly differs from the 
one mentioned 
in Section \ref{s:Three}. From Eq. \ref{e2} we obtain
\begin{equation}\label{e7}
\frac{SFR}{m}= \widetilde{A}\widetilde{f}(t)+\widetilde{B}\Delta,
\end{equation}
The values of $\widetilde{A}_i$ and $\widetilde{B}_i$ are obtained by the
least squares method using the approximation in Eq.~\ref{e7}. They slightly 
differ from 
the values $A_i$ and $B_i$ used in Section \ref{s:Three}.
Then $m$ and $T$ are derived from minimization of the expression
\begin{equation}\label{e8}
\widetilde{U}=\sum_{i=1}^3\left(\frac{SFR_i/m-\widetilde{A}_i\widetilde{f}_i(t)-
\widetilde{B}_i\Delta}{\widetilde{\sigma}_i}\right)^2.
\end{equation}
Here $\widetilde{\sigma}_i$ are mean dispersions of the data 
around the approximation in Eq. \ref{e7} at three wavelengths.
The condition $\partial\widetilde{U}/\partial T=0$ corresponds 
to Eq. \ref{e6}, but with tildes above the $A$, $B$, $f$ and $p$.
The condition
$\partial\widetilde{U}/\partial m=0$ yields
\begin{equation}\label{e9}
\displaystyle m=\frac{\sum_{i=1}^3 SFR_i^2\widetilde{\sigma}_i^{-2}}
{\sum_{i=1}^3 SFR_i(\widetilde{A}_i\widetilde{f}_i+\widetilde{B}_i\Delta)\widetilde{\sigma}_i^{-2}}.
\end{equation}
instead of Eq. \ref{e5}.
If the numerical solution of Eq. \ref{e6} with Eq. \ref{e9}
and $\widetilde{f}_i=\exp (-\widetilde{p}_it)$ corresponds to $t>0$, we have the
estimation $\widetilde{T}_e=T_0+t$ with the different value of $t$
as compared to the one in Section \ref{s:Three}. In the case
$t<0$ we derive the estimation $\widetilde {T}_e\le T_0$ while Eq.
\ref{e9} with $\widetilde{f}_i=1$ provides the value of $\widetilde{m}_e$.
As a result we obtain somewhat different estimations of the ages of
the starburst and the masses of the young stellar population for
LCGs. We denote the estimation obtained by the prior method as
$m_e$ and $T_e$ and the estimation derived by the latter method as
$\widetilde{m}_e$ and $\widetilde{T}_e$.

It is not known a priori which of these methods is better, but 
the detailed analysis carried out in
Section \ref{s:Precision} shows that the alternative
version of the method provides much worse estimation than the
first one. This follows from the comparison of the values obtained
by both methods and the values $m_f$ and $T_f$ derived from SED fitting. 
Hereafter the subscript $f$ means ``fitted'' and refers to the values 
obtained by \citet{I11}.

\section{Sample and subsamples}\label{s:Sample}

For the determination of the starburst age $T$ and the mass $m$ of the young
stellar population in the individual galaxy we need SFR$_i$ derived from the
H$\alpha$, FUV, and NUV luminosities and $\Delta$ derived from the
oxygen abundance [O]. We use the calibration sample 
of about 800 LCGs constructed by \citet{I11}. 
The FUV and NUV fluxes for LCGs
were extracted from the {\sl GALEX} Medium Imaging Survey (MIS) and
All-sky Imaging Survey (AIS) data. These data combined
with the SDSS data on H$\alpha$ fluxes and redshifts \citep{I11}
provide the determination
of the galaxies H$\alpha$ and UV luminosities.

Because the radiation of galaxies is reduced by dust extinction, we applied
reddening corrections to H$\alpha$ and UV band fluxes using \citet{C89}
reddening law. All H$\alpha$ emission line luminosities are also corrected for 
aperture \citep[see details in ][]{PII}.

In \citet{PII} the sample of LCGs was splitted into two subsamples of ``regular'' galaxies
with the round shape and ``irregular'' galaxies with some sign of disturbed
morphology suggesting the presence of two or more star-forming regions and
their interaction.
To clarify the impact of the LCG's morphology
on the obtained parameters we instead of these two subsamples
produce seven subsamples on the base of detailed visual
examination of galaxies morphology on SDSS images. 
We marked the galaxies with 2-3 evident knots of star formation separated by 
$\le 10$\arcsec\ in the
subsample of ``irregular'' galaxies. In addition we marked galaxies
with a single star-forming region, but with an obvious visual
asymmetry, e.g. in the form of some small ``tail'' or a sign of a
``tail''. Summarising, we split the LCG sample on the following
subsamples. I -- 
``regular'' galaxies. II --  ``irregular'' galaxies. These
subsamples are practically the same as the subsamples used by
papers \citet{I11} and \citet{PII}. III --  ``irregular''
galaxies with a single star-forming region. It is formed
from the subsample II after discarding of galaxies with two or
three possible star-forming regions. IV --  ``irregular''
galaxies with a single star-forming region without obvious visual
asymmetry. IV is a subset of III and III is a subset of II.
Combining I with II, III or IV we produce three more subsamples. V is
a combination of subsamples I and II, this is the entire initial
sample. VI is a combination of subsamples I and III, it contains
all the galaxies with a single star-forming region. VII is a
combination of subsamples I and IV, it contains all galaxies with
a single star-forming region without obvious asymmetry.

These subsamples can be additionally constrained according to the errors in 
the measured UV fluxes. Our initial entire LCG sample includes only galaxies
with the UV flux errors not exceeding 50\%,
but we can also impose a more strict limitation
on the errors. The flux error limitation in the first column of Table \ref{t1}
is shown as the percentage of the UV flux. The word ``all'' means
the initial 50\% error level, designations 30\%, 20\% and 10\% indicate the 
reduced thresholds of UV flux errors. Note that these thresholds apply
to the fluxes both in FUV and NUV ranges.

\begin{table*}[tb]
\small
\caption{Values of the coefficients in Eqs. \ref{e2} and \ref{e3}
for different subsamples of LCGs}
\begin{tabular}{@{}l@{ }c@{ }c@{ }c@{ }c@{ }c@{ }c@{ }c@{ }c@{ }c@{ }c@{ }c@{ }c@{ }c@{ }c@{ }c@{}}
\hline
\multirow{2}{*}{Err}&\multicolumn{5}{c}{H$\alpha$}&\multicolumn{5}{c}{FUV}&\multicolumn{5}{c}{NUV}\\
\cline{2-16}
&$N$&$\sigma_1$&$p_1$&$\bigstrut A_1\times 10^{7}$&$B_1\times 10^{8}$
&$N$&$\sigma_2$&$p_2$&$\bigstrut A_2\times 10^{7}$&$B_2\times 10^{8}$
&$N$&$\sigma_3$&$p_3$&$\bigstrut A_3\times 10^{7}$&$B_3\times 10^{8}$\\ \hline
\multicolumn{16}{c}{I. ``Regular''} \\ \hline
all&$276$&$3.3$&$0.67$&$1.92\pm0.03$&$-4.4\pm1.0$&$212$&$4.4$&$0.39$&$1.33\pm0.04$&$-5.6\pm1.4
$&$232$&$5.1$&$0.31$&$1.58\pm0.04$&$-5.1\pm1.5$\\
30\%&$172$&$3.4$&$0.72$&$1.97\pm0.04$&$-5.0\pm1.2$&$169$&$4.4$&$0.39$&$1.36\pm0.04$&$-7.4\pm1.5
$&$169$&$4.9$&$0.32$&$1.61\pm0.04$&$-4.8\pm1.7$\\
20\%&$120$&$3.1$&$0.70$&$1.98\pm0.05$&$-3.1\pm1.2$&$117$&$3.7$&$0.21$&$0.98\pm0.03$&$-9.2\pm1.5
$&$117$&$3.6$&$0.13$&$1.14\pm0.03$&$-5.1\pm1.7$\\
10\%&$47$&$2.5$&$0.73$&$2.32\pm0.09$&$-7.4\pm2.5$&$45$&$2.3$&$0.47$&$1.18\pm0.07$&$-12.6\pm2.4
$&$45$&$2.2$&$0.30$&$1.32\pm0.06$&$-10.7\pm2.4$\\ \hline
\multicolumn{16}{c}{II. ``Irregular''} \\ \hline
all&$519$&$3.6$&$0.64$&$1.82\pm0.02$&$-3.3\pm0.6$&$418$&$4.4$&$0.40$&$0.98\pm0.02$&$-3.9\pm0.9
$&$435$&$7.3$&$0.33$&$1.64\pm0.03$&$-12.6\pm1.5$\\
30\%&$391$&$3.3$&$0.67$&$1.90\pm0.02$&$-3.8\pm0.7$&$388$&$4.4$&$0.41$&$1.00\pm0.02$&$-3.6\pm0.9
$&$387$&$6.3$&$0.33$&$1.60\pm0.03$&$-11.9\pm1.3$\\
20\%&$348$&$3.3$&$0.66$&$1.90\pm0.03$&$-5.7\pm0.9$&$346$&$4.3$&$0.41$&$1.03\pm0.03$&$-6.4\pm1.2
$&$345$&$6.0$&$0.34$&$1.59\pm0.04$&$-13.1\pm1.7$\\
10\%&$171$&$3.5$&$0.64$&$1.80\pm0.04$&$-5.1\pm1.2$&$170$&$4.7$&$0.42$&$1.01\pm0.04$&$-4.9\pm1.6
$&$169$&$6.3$&$0.34$&$1.61\pm0.04$&$-14.8\pm2.1$\\ \hline
\multicolumn{16}{c}{III. ``Irregular'' with a single star-forming region} \\ \hline
all&$415$&$3.4$&$0.65$&$1.85\pm0.02$&$-2.6\pm0.7$&$333$&$3.8$&$0.48$&$1.07\pm0.03$&$-2.2\pm0.9
$&$351$&$7.2$&$0.39$&$1.79\pm0.04$&$-13.9\pm1.6$\\
30\%&$312$&$2.8$&$0.73$&$2.07\pm0.03$&$-4.0\pm0.7$&$312$&$3.7$&$0.48$&$1.07\pm0.03$&$-2.3\pm0.9
$&$312$&$6.1$&$0.37$&$1.69\pm0.04$&$-12.9\pm1.4$\\
20\%&$276$&$2.8$&$0.73$&$2.11\pm0.03$&$-5.9\pm0.9$&$276$&$3.7$&$0.48$&$1.09\pm0.03$&$-4.8\pm1.2
$&$276$&$6.0$&$0.40$&$1.70\pm0.04$&$-15.5\pm1.8$\\
10\%&$127$&$2.2$&$0.74$&$2.11\pm0.03$&$-5.4\pm0.9$&$127$&$3.7$&$0.51$&$1.11\pm0.04$&$-3.3\pm1.5
$&$127$&$5.3$&$0.42$&$1.77\pm0.05$&$-19.7\pm2.1$\\ \hline
\multicolumn{16}{c}{IV. ``Irregular'' with a single star-forming region without asymmetry} \\ \hline
all&$275$&$3.5$&$0.70$&$2.01\pm0.03$&$-4.2\pm0.9$&$219$&$3.9$&$0.42$&$1.01\pm0.04$&$-0.3\pm1.0
$&$232$&$7.2$&$0.36$&$1.72\pm0.05$&$-14.3\pm1.9$\\
30\%&$202$&$2.9$&$0.73$&$2.06\pm0.03$&$-3.6\pm0.8$&$202$&$3.9$&$0.41$&$1.00\pm0.03$&$-0.3\pm1.1
$&$202$&$5.3$&$0.33$&$1.58\pm0.03$&$-13.3\pm1.4$\\
20\%&$175$&$2.9$&$0.74$&$2.13\pm0.03$&$-6.3\pm1.1$&$175$&$3.9$&$0.42$&$1.03\pm0.03$&$-2.7\pm1.4
$&$175$&$5.2$&$0.33$&$1.57\pm0.04$&$-15.7\pm1.9$\\
10\%&$77$&$1.7$&$0.74$&$2.12\pm0.03$&$-5.0\pm0.9$&$77$&$3.8$&$0.50$&$1.13\pm0.09$&$\phantom{-}1.9\pm2.1
$&$77$&$5.2$&$0.41$&$1.85\pm0.06$&$-14.0\pm4.3$\\ \hline
\multicolumn{16}{c}{V=I+II. All} \\ \hline
all&$795$&$3.5$&$0.65$&$1.85\pm0.02$&$-3.6\pm0.5$&$630$&$4.6$&$0.43$&$1.12\pm0.02$&$-5.1\pm0.8
$&$667$&$6.7$&$0.33$&$1.63\pm0.03$&$-10.2\pm1.1$\\
30\%&$563$&$3.3$&$0.68$&$1.91\pm0.02$&$-4.1\pm0.6$&$557$&$4.7$&$0.43$&$1.14\pm0.02$&$-5.2\pm0.8
$&$556$&$6.0$&$0.33$&$1.60\pm0.02$&$-10.0\pm1.1$\\
20\%&$468$&$3.3$&$0.67$&$1.91\pm0.02$&$-4.8\pm0.7$&$463$&$4.4$&$0.35$&$1.03\pm0.02$&$-8.3\pm1.0
$&$462$&$5.6$&$0.30$&$1.49\pm0.03$&$-10.8\pm1.2$\\
10\%&$218$&$3.4$&$0.66$&$1.87\pm0.03$&$-5.8\pm1.1$&$215$&$4.3$&$0.43$&$1.04\pm0.03$&$-5.7\pm1.4
$&$214$&$5.7$&$0.33$&$1.57\pm0.04$&$-14.2\pm1.8$\\ \hline
\multicolumn{16}{c}{VI=I+III. All with a single star-forming region} \\ \hline
all&$691$&$3.4$&$0.66$&$1.88\pm0.02$&$-3.2\pm0.5$&$545$&$4.3$&$0.47$&$1.20\pm0.02$&$-3.9\pm0.8
$&$583$&$6.5$&$0.37$&$1.72\pm0.03$&$-10.8\pm1.2$\\
30\%&$484$&$3.0$&$0.72$&$2.02\pm0.02$&$-4.2\pm0.6$&$481$&$4.4$&$0.46$&$1.19\pm0.02$&$-4.3\pm0.8
$&$481$&$5.7$&$0.36$&$1.67\pm0.03$&$-10.5\pm1.1$\\
20\%&$396$&$2.9$&$0.72$&$2.06\pm0.02$&$-4.8\pm0.7$&$393$&$4.0$&$0.39$&$1.07\pm0.02$&$-7.7\pm1.0
$&$393$&$5.3$&$0.34$&$1.55\pm0.03$&$-12.1\pm1.3$\\
10\%&$173$&$2.3$&$0.75$&$2.16\pm0.03$&$-5.7\pm0.9$&$172$&$3.5$&$0.51$&$1.15\pm0.03$&$-4.5\pm1.3
$&$172$&$4.8$&$0.39$&$1.67\pm0.04$&$-18.3\pm1.8$\\ \hline
\multicolumn{16}{c}{VII=I+IV. All with a single star-forming region without asymmetry} \\ \hline
all&$551$&$3.4$&$0.69$&$1.98\pm0.02$&$-4.4\pm0.7$&$431$&$4.5$&$0.43$&$1.18\pm0.02$&$-2.3\pm0.9
$&$464$&$6.3$&$0.33$&$1.64\pm0.03$&$-10.0\pm1.2$\\
30\%&$374$&$3.1$&$0.73$&$2.03\pm0.03$&$-4.2\pm0.7$&$371$&$4.6$&$0.43$&$1.19\pm0.03$&$-2.7\pm1.0
$&$371$&$5.2$&$0.33$&$1.61\pm0.03$&$-9.9\pm1.1$\\
20\%&$295$&$3.0$&$0.72$&$2.06\pm0.03$&$-4.7\pm0.8$&$292$&$4.2$&$0.35$&$1.04\pm0.02$&$-6.2\pm1.1
$&$292$&$4.7$&$0.27$&$1.43\pm0.02$&$-9.2\pm1.3$\\
10\%&$124$&$2.1$&$0.76$&$2.22\pm0.03$&$-5.4\pm1.0$&$122$&$3.5$&$0.52$&$1.24\pm0.04$&$-1.5\pm1.7
$&$122$&$4.5$&$0.36$&$1.67\pm0.04$&$-13.6\pm2.1$\\ \hline

\end{tabular}\label{t1}

{\sc Note.} The values of $p_i$ are in
Myr$^{-1}$, standard deviations $\sigma_i$ are in $M_{\odot}$ yr$^{-1}$, coefficients $A_i$ and
$B_i$ in Eq. \ref{e2} are in yr$^{-1}$, $N$ is the number of the galaxies in the
subsample. The ``Err'' column indicates the threshold of the
measured UV flux errors, the label ``all'' corresponds to the 50\%
level.
\end{table*}

\section{Determination of parameters $A, B, p$ and $\sigma$}\label{s:Parameters}

Three sets of parameters $A_i, B_i, p_i, 
\sigma_i$ in approximations Eq. \ref{e2} are needed to estimate $m$ and $T$. 
The determination of these parameters is
described in details by \citet{PII}. It was carried out by the
least squares method and the linear regression analysis.

We use the least squares method to calculate the sets of parameters 
in Eq. \ref{e2} for all subsamples described in Section \ref{s:Sample}.
The number $N$ of the galaxies, the values of $p_i$ in Myr$^{-1}$, 
$\sigma_i$ in $M_{\odot}$ yr$^{-1}$, $A_i$ and $B_i$ in yr$^{-1}$ are shown 
in Table \ref{t1}.

The parameters in \citet{PII} are almost the same as
the ones in the first
row (label ``all'' in the first column) in the part of Table \ref{t1} 
corresponding to the subsample V.
It differs from the sample used in \citet{PII} only by
excluding one obvious outlier from the subsample of ``regular'' galaxies.

One can see from Table \ref{t1} that discarding galaxies with possible several 
star-forming regions and asymmetry noticeably decreases the variance, as well 
as the flux errors limitation. This effect becomes stronger if the galaxies 
with the pronounced asymmetry are also excluded. The F-test 
\citep{F,H} shows that the statistical significance of the first term on the 
right side of Eq. \ref{e2} is very high and the statistical significance of 
the second term exceeds 99\% level for the majority of subsamples. It becomes 
statistically insignificant only for FUV radiation and subsamples IV
and VII with flux errors less than 10\%.

The values of $\sigma_1$ are less that $\sigma_2$ and $\sigma_3$ due to a 
better accuracy of H$\alpha$ flux measurements. 
The values of $\sigma_2$ and $\sigma_3$ for the samples with flux errors 
limitation are decreased with increasing of flux accuracy.
The difference between $A_i$ values indicates that the $k$ factors have to be 
corrected for better agreement between SFRs derived from fluxes at different
wavelengths. Corresponding changes were proposed and the ratios of new $k$ values 
for FUV, NUV and H$\alpha$ were found in \citep{PII}. In the present paper we don't 
compare the SFRs derived using different tracers. 
We use the equal $k$ values for FUV and NUV ranges and use the $k$ value for 
H$\alpha$ line estimated for the galaxies with different properties \citep{K98}.
This do not affect the accuracy of $m$ and $T$ estimation but leads to the difference
between the values of $A_i$ for different tracers.

The values of the coefficients $A_i, B_i$ and $p_i$, derived
from different subsamples, vary insignificantly. Nevertheless,
small variations of these coefficients result in a noticeable
change of data scatter and $\sigma_i$. Thus, it is important to
use a set of the coefficients calculated for a properly refined
subsample.

We choose the values for the subsample VII with UV flux errors smaller 
than 20\% as the set of
coefficients for Eq. \ref{e2}. For each galaxy we estimate $T_e$ and $m_e$ using the
coefficients $\sigma_i=(3.0, 4.2, 4.7)$, $p_i=(0.72, 0.35, 0.27)$, $A_i=(2.06\times 10^{-7},
1.04\times 10^{-7}, 1.43\times 10^{-7})$ and $B_i=(-4.72\times 10^{-8},-6.21\times 10^{-8},
-9.21\times 10^{-8})$.

The parameters $\widetilde{p}_i$, 
$\widetilde{A}_i$, $\widetilde{B}_i$ and $\widetilde{\sigma}_i$ in 
Eq. \ref{e7} for the subsamples shown in Table \ref{t1} were 
calculated by the least squares method as well. 
Note that this approximation is the same as given 
by Eq. \ref{e2}, but with statistical weights proportional to $m^{-2}$, which 
makes it more sensitive to the data for small galaxies.
The values of $\widetilde{p}$ in this case are larger than those for
the approximation Eq. \ref{e2} and closer to each other. We choose the values 
for the subsample
VII with UV flux errors smaller than 30\% as the sets of coefficients for 
Eq. \ref{e7}. We estimate $\widetilde{T}_e$ and $\widetilde{m}_e$ using the 
coefficients $\widetilde{\sigma}_i=(6.0 \times 10^{-8},
6.1 \times 10^{-8}, 8.1 \times 10^{-8})$ yr$^{-1}$, $\widetilde{p}_i=(0.70, 0.51, 0.50)$,
$\widetilde{A}_i=(2.20\times 10^{-7},1.26\times 10^{-7},1.65\times 10^{-7})$ and
$\widetilde{B}_i=(-6.90\times 10^{-8},-9.72\times 10^{-8},-8.93\times 10^{-8})$. The
parameters for other subsamples are close to these values.

\section{On the method accuracy}\label{s:Precision}

Compare the values $m_e$ and $T_e$ obtained by the method 
proposed in Section \ref{s:Three} and based on Eqs. \ref{e5}-\ref{e6}
with the values $m_f$ and $T_f$ obtained in \citet{I11} by fitting
the galaxy SEDs. Although we can estimate $m_e$ and $T_e$ for almost every 
galaxy from the sample, using the entire sample is a poor choice for such a 
comparison due to the presence of galaxies with several star-forming regions 
for which the values of $m_e$ and $T_e$ can differ substantially. The described
method is intended for quick estimation of the values $m_e$ and $T_e$ in large 
LCGs samples without SED fitting.
Note that such samples may contain both ``regular'' and ``irregular'' galaxies. We can exclude
galaxies with several star-forming regions and, if desired, galaxies with an asymmetry by
visual inspection of SDSS images. Thus we use the subsamples VI or VII for more reliable
and accurate comparison. In addition we can constrain the values of UV flux errors.

\begin{figure*}[tb]
\includegraphics[width=\columnwidth]{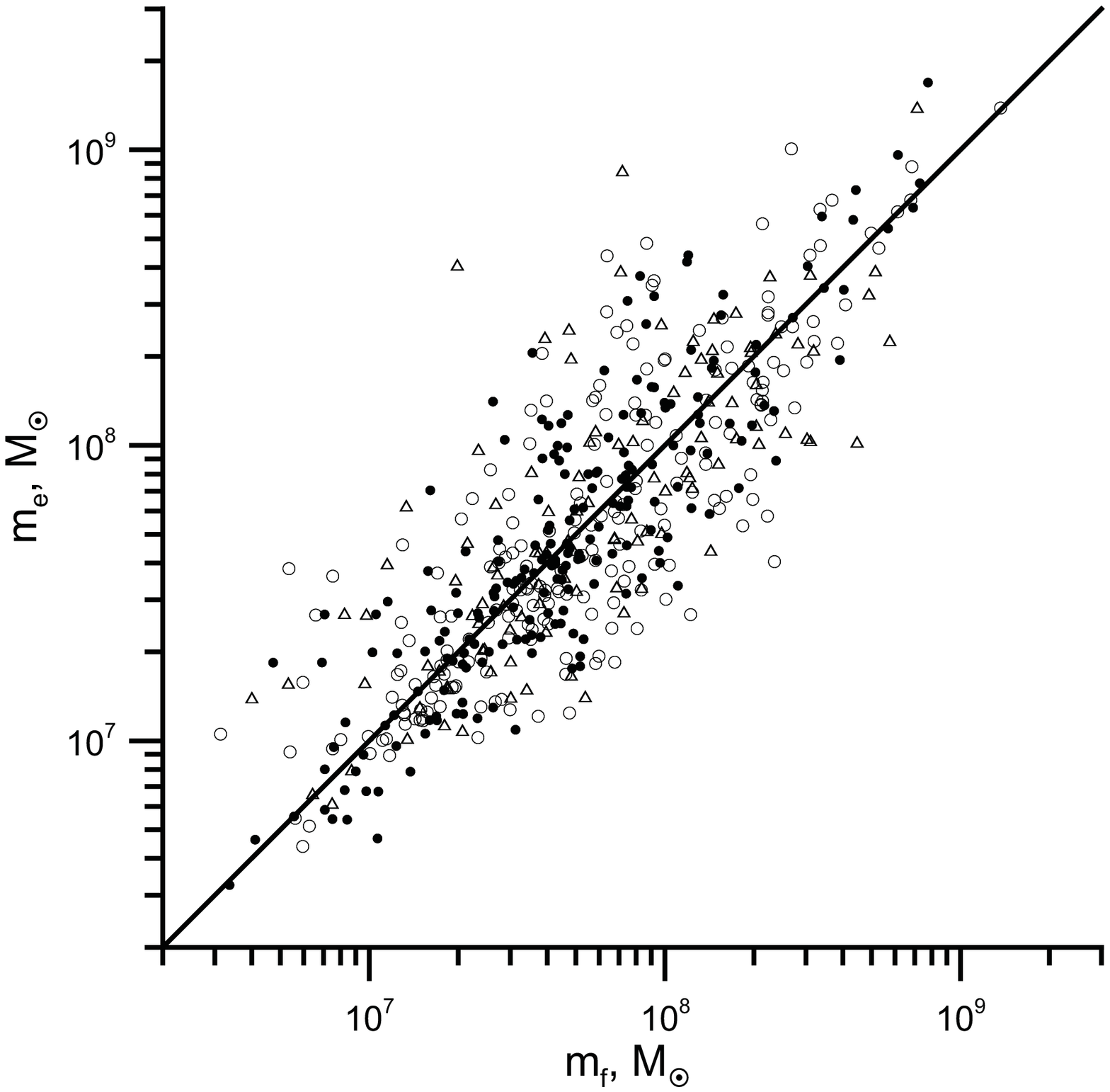}
\includegraphics[width=\columnwidth]{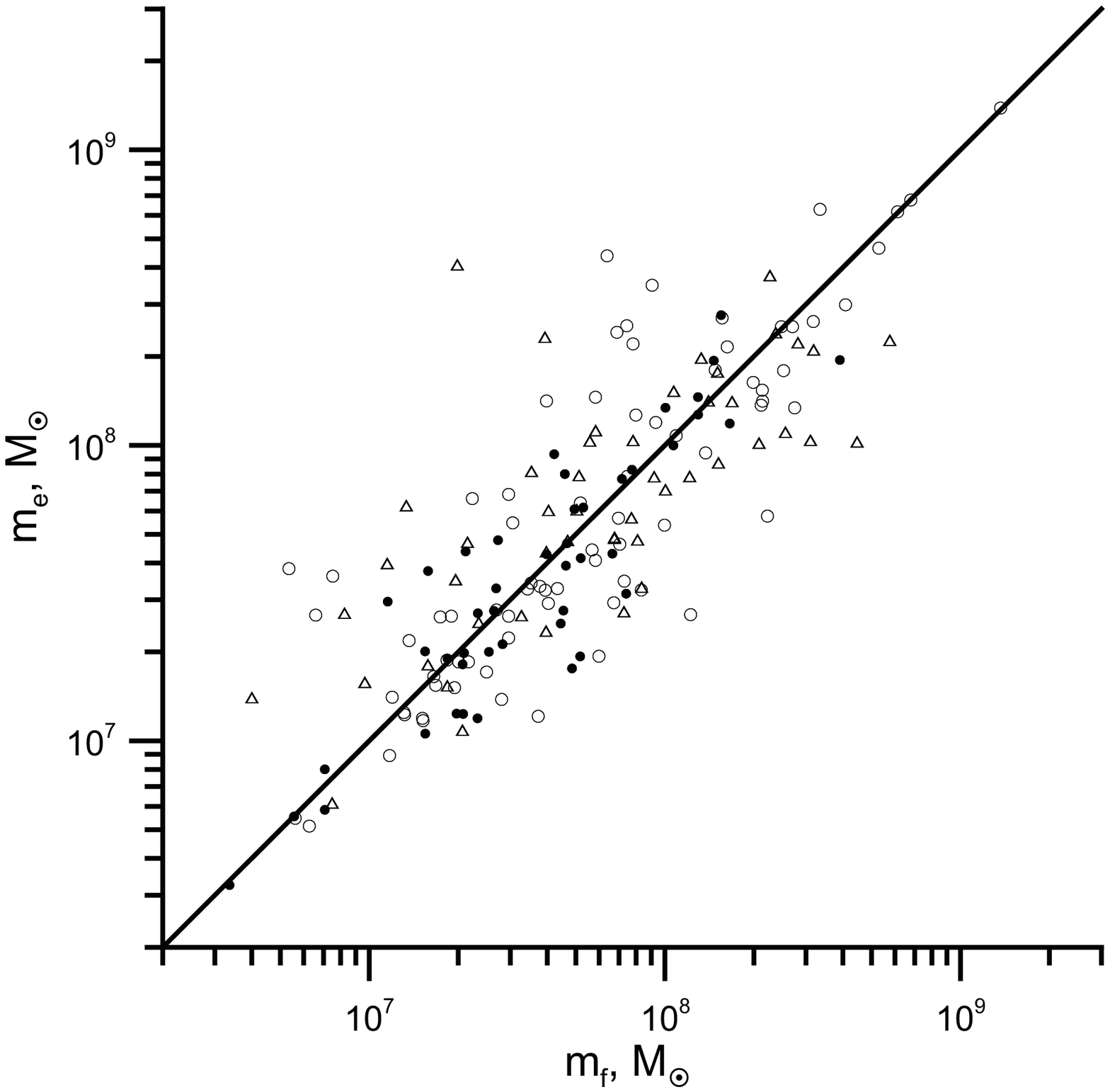}
\caption{Comparison of the masses $m_e$ of young stellar
population derived from the estimation proposed in Section \ref{s:Three} and
masses $m_f$ derived from the SED fitting by \citet{I11}. Dots, open circles 
and triangles correspond to subsamples of galaxies with ``regular'',
``irregular'' symmetric and ``irregular'' asymmetric shape. Left
panel shows all galaxies from subsample VI, right panel
shows only the galaxies with flux errors in UV ranges less than 10\%. 
Solid line corresponds to the equal values}
\label{f1}
\end{figure*}

In Fig. \ref{f1} we show an example for the subsample VI.
Dots correspond to
``regular'' galaxies from subsample I, open circles -- to ``irregular'' galaxies with a single star-forming
region without obvious asymmetry from subsample IV, and triangles -- to ``irregular'' galaxies with a
single star-forming region with obvious asymmetry. Triangles together with open circles
correspond to subsample III. It is seen that the estimated $m$ roughly match 
the masses obtained from SED fitting. The largest deviations occur for 
asymmetric LCGs.

We estimate the accuracy of the proposed method. Note that three values SFR$_i$
are used to derive two values $m_e$ and $T_e$. Thus, we still have a
possibility to obtain one additional parameter for each LCG which is
a characteristic of the consistency between different $m$ estimates.
Given the value of $T_e$ derived from Eqs. \ref{e5}-\ref{e6} we can obtain 
three different estimations of $m_e$ using only galaxy luminosities at three 
wavelengths as
\begin{equation}\label{e10}
m_i=\frac{SFR_i}{A_if_i+B_i\Delta}
\end{equation}
The estimations in Eqs. \ref{e5} or \ref{e9} are weighted averages of these 
values. 

We choose the root-mean-square value of the logarithm of the $m_e$ to $m_f$ 
ratio as the indicator of deviations and define the quantity
\begin{equation}\label{ea}
q=\left(\frac{\sum_{k=1}^N\log^2(m_e/m_f)_k}{N}\right)^{1/2}.
\end{equation}
In Table \ref{t3} we present the values of $q$
for different subsamples.

\begin{table}[tb]
\caption{Values of $q$ from Eq. \ref{ea} for different
subsamples derived from luminosities at three 
wavelength ranges and the oxygen abundance data}
\begin{tabular}{@{}l@{ }l@{ }l@{ }l@{ }l@{}}
\hline
      &\multicolumn{4}{c}{Flux error threshold} \\ \cline{2-5}
Sample&all (50\%)&30\%&20\%&10\%\\ \hline
I.&$0.251(207)$&$0.245(169)$&$0.210(117)$&$0.201(45)$\\
II.&$0.343(408)$&$0.344(386)$&$0.350(344)$&$0.388(168)$\\
III.&$0.305(329)$&$0.303(311)$&$0.305(275)$&$0.322(126)$\\
IV.&$0.290(218)$&$0.285(202)$&$0.289(175)$&$0.300(77)$\\
V.&$0.315(615)$&$0.318(555)$&$0.321(461)$&$0.358(213)$\\
VI.&$0.285(536)$&$0.284(480)$&$0.280(392)$&$0.295(171)$\\
VII.&$0.271(425)$&$0.267(371)$&$0.260(292)$&$0.267(122)$\\ \hline
\end{tabular}
{\sc Note}. The number of galaxies is shown in brackets.
\label{t3}
\end{table}

One can see from Table \ref{t3} that the values of $q$ are substantially 
larger for the II and V subsamples containing galaxies with several 
star-forming regions as compared to the typical $q$ values of about 0.27. 
The typical ratio of $m_e$ to $m_f$ 
(or $m_f$ to $m_e$) is
$10^{0.27}\approx 1.86$. Thus, the typical values 
of the mass of young stellar population obtained from fitting of $m_f$ fall in 
the range from $1.86 m_e$ to $m_e/1.86=0.54m_e$, where $m_e$ is the estimation 
obtained from the proposed method. Therefore, we can estimate $m$ with a 
relative 
error up to a factor of 1.86. Note that for different galaxies a typical value 
of $m$ can vary within wide limits and differs by 3 orders of magnitude.

\begin{figure}[tb]
\includegraphics[width=\columnwidth]{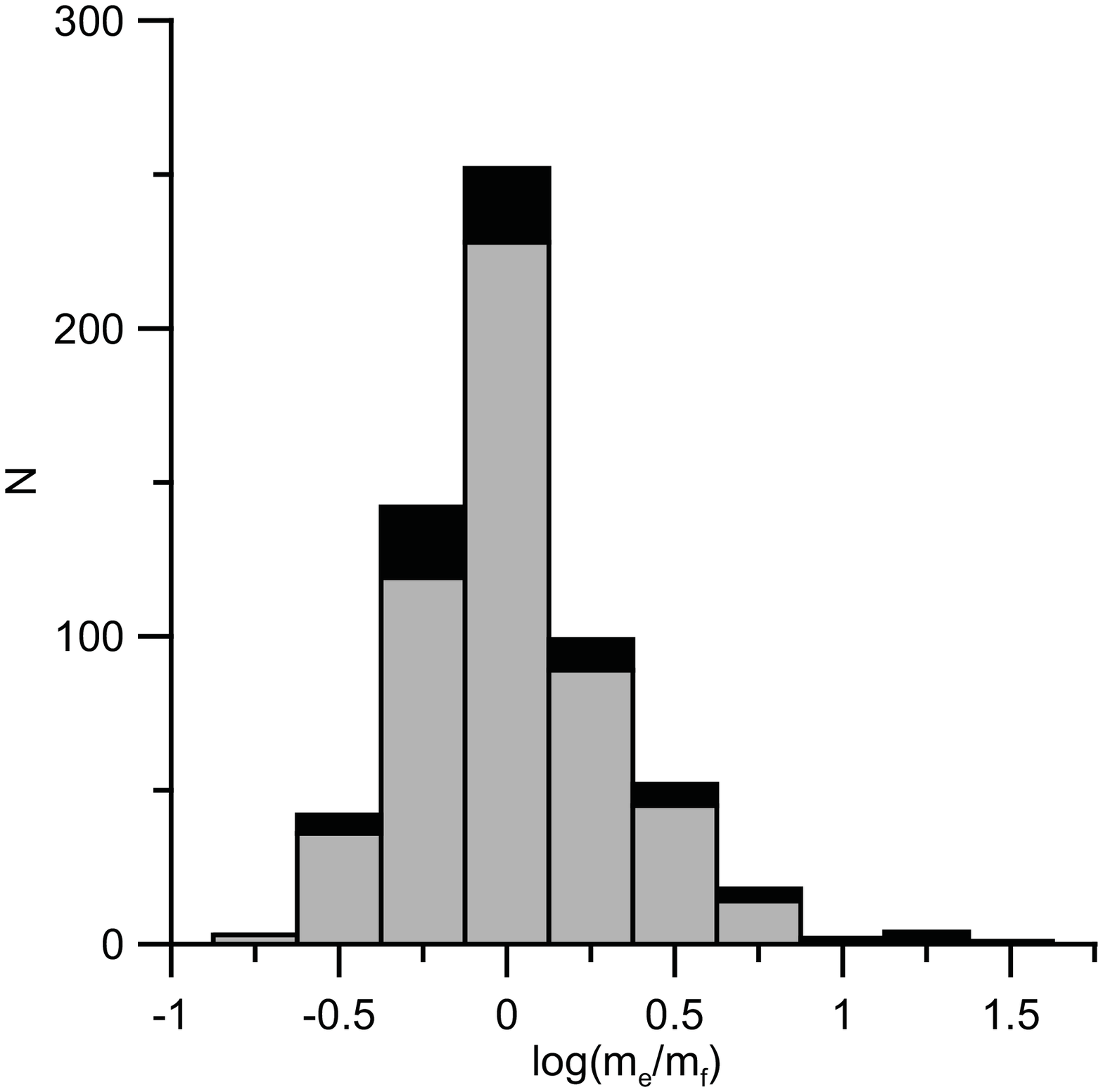}
\caption{The distribution of the logarithm of 
the ratio $m_e$ of young stellar
population derived from the estimation proposed in Section \ref{s:Three} to
masses $m_f$ derived from the SED fitting by \citet{I11} for the whole sample. 
Grey-shaded histogram corresponds to the galaxies with a single 
star-forming region (subsample VI)
and black-shaded histogram corresponds to the galaxies with obvious several 
star-forming regions. The whole bar (grey plus black) corresponds to the whole sample}
\label{f1a}
\end{figure}

In Fig. \ref{f1a} we show the distribution of the logarithm of 
the ratio $m_e$ to $m_f$ for the whole sample (sample V). We plot the 
stacked bar chart 
for the galaxies with a single star-forming region (subsample VI,
grey-shaded histogram) and for the galaxies with several star-forming regions 
(galaxies entering in the subsample V and missed the subsample VI, 
black-shaded histogram).
One can see that the proposed technique provides better estimation for the 
galaxies with a single star-forming regions and it should not be applied to 
galaxies with several obvious knots of star formation.

To estimate the errors in estimation of $T_e$ values
we consider  $T_e$ and $T_f$ for the subsample VII containing 425 galaxies. 
We know $T_f$ for each galaxy, as well as either the value of $T_e$ 
if $T_e>T_0$ or an upper limit $T_0$ otherwise. The latter circumstance puts 
some limitations in the direct comparison of $T_e$ 
and $T_f$ values. We can directly compare $T_e$ and $T_f$ only for 178 
galaxies with $T_e>T_0$ and $T_f>T_0$. 
The mean-root-square value of the difference is 
0.87 Myr. For 95 galaxies with $T_e\le T_0$ and $T_f\le T_0$ we can only 
acknowledge a matching of the conclusions from SED fitting and 
our estimation. For 68 galaxies with $T_e\le T_0$ and $T_f>T_0$ we for 
convenience could assign the value $T_e = T_0=3.2$ Myr. This keeps the 
time intervals between 
starburst age $T_f$ and the upper end of the possible interval $0<T_e\le T_0$. 
The mean-root-square value of the difference $T_f-T_e$ in this case is equal 
to 0.91 Myr. It is quite reasonable to consider the remaining 84 galaxies with 
$T_e> T_0$ and $T_f\le T_0$ in the same way, namely assign the value 
$T_f = T_0=3.2$ Myr instead of the value from  SED fitting. The 
mean-root-square value of the difference $T_f-T_e$ in this case is equal 
to 1.3 Myr. 

\begin{figure}[tb]
\includegraphics[width=\columnwidth]{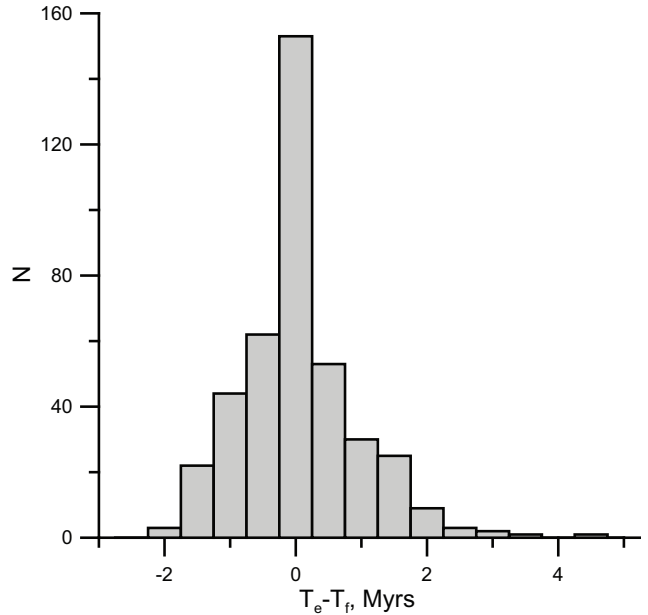}
\caption{The distribution of the difference between the starburst 
ages $T_e$ estimated from
the proposed method and $T_f$ derived from the SED fitting by \citet{I11} f
or the subsample VII}
\label{f2}
\end{figure}

In Fig. \ref{f2} a histogram of the difference between the 
estimated $T_e$ and fitted $T_f$ is shown for the entire 
subsample VII. Here we substituted $T_e=T_0$ if $T_e\le T_0$ and/or $T_f=T_0$ 
if $T_f \le T_0$. A typical value of the difference $T_f-T_e$ is less than 1 
Myr. Its overall mean-root-square value is 0.87 Myr. There is no systematic 
shift of the starburst age, as one can see in Fig. \ref{f2}.

\begin{figure}[tb]
\includegraphics[width=\columnwidth]{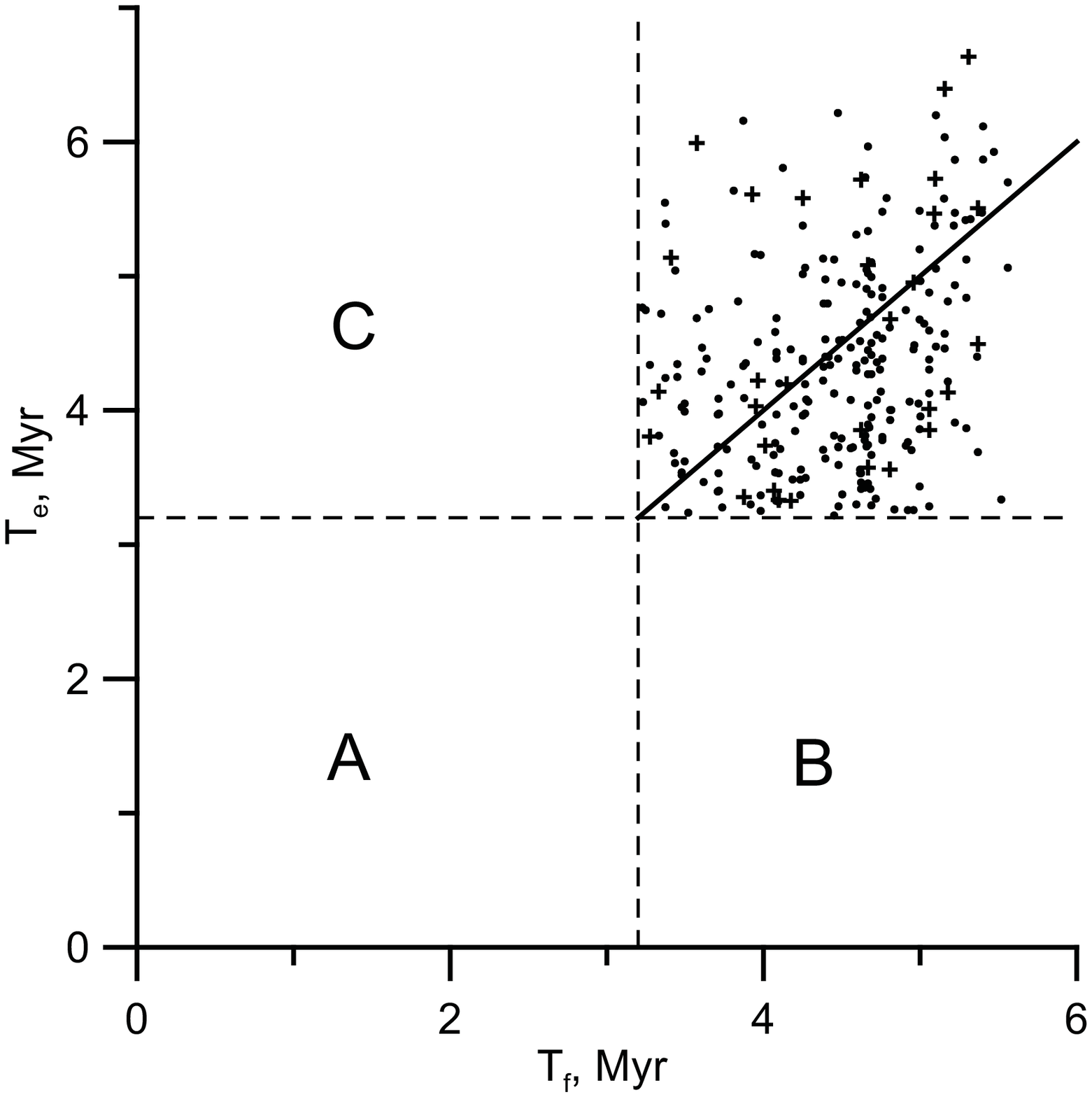}
\caption{Comparison of the starburst ages $T_e$ derived from the estimation 
proposed 
in Section \ref{s:Three} and $T_f$ derived from the SED fitting by \citet{I11}. Dots
and crosses correspond to galaxies with a single star-forming region and
with obvious several star-forming regions, respectively. Solid line 
is the line of equal ages} 
\label{f4}
\end{figure}

In Fig. \ref{f4} we plot $T_e$ vs. $T_f$ for the whole 
sample. Their comparison makes sense only if both $T_e$ and $T_f$ are
greater than $T_0=3.2$ Myr. In the Figure, 235 galaxies with a single 
star-forming region (subsample VI) are shown by dots. 
We discard one outlier from this subsample. 
Thirty galaxies with several star-forming regions are shown by crosses. 
The solid line is the line of equal ages.
 
\begin{figure}[tb]
\includegraphics[width=\columnwidth]{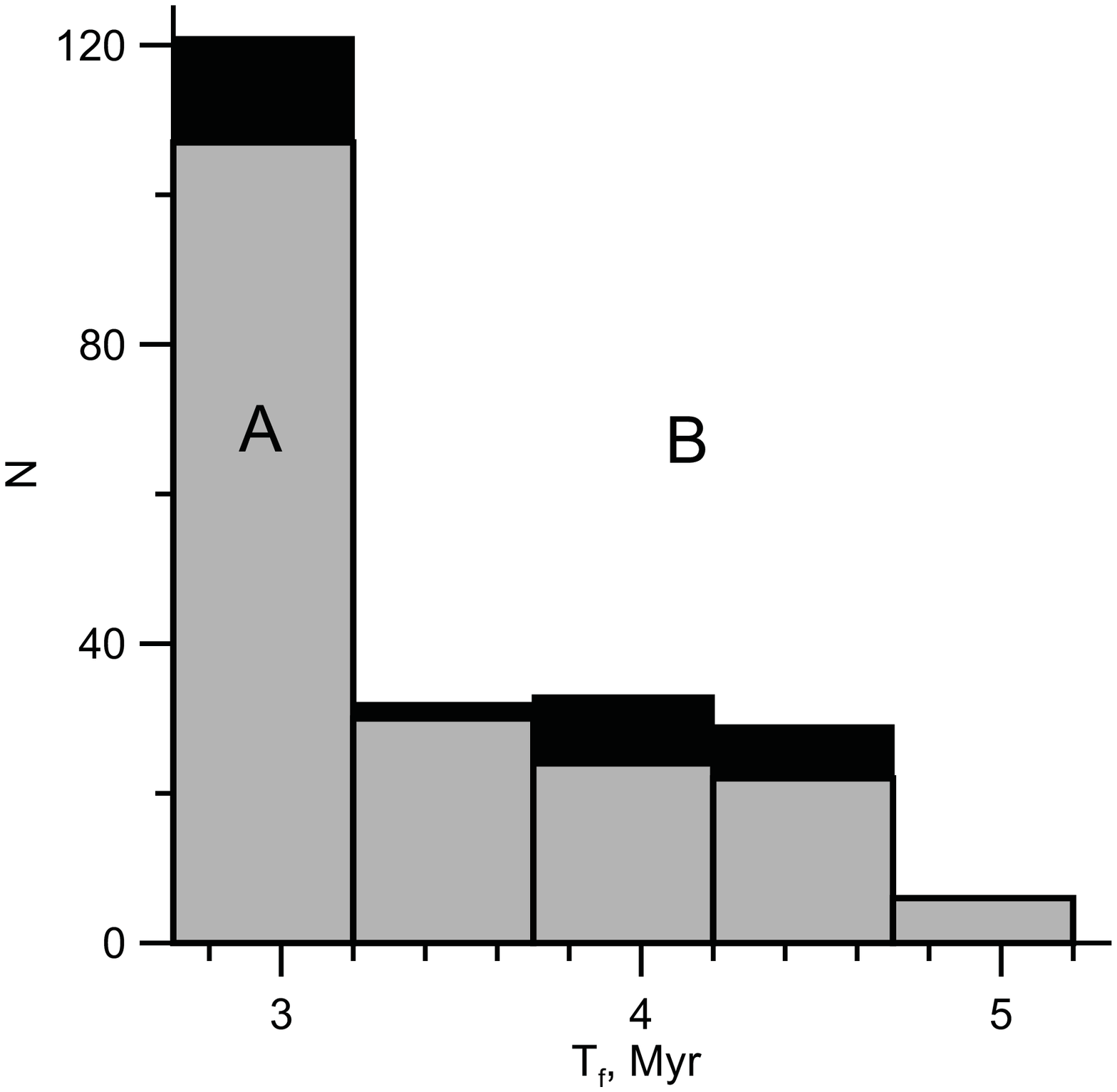}
\caption{The $T_f$ distribution in the case $T_e\le T_0$
(corresponding to zones A and B in
Fig. \ref{f4}). Grey-shaded histogram corresponds to the subsample VI of 
galaxies 
with a single star-forming region and black-shaded histogram to the galaxies 
with several star-forming regions. The whole bar (grey plus black) corresponds to the whole sample}
\label{f5}
\end{figure}

We divide the rest of the plot area into three zones A, B and C. Zone A, 
corresponding to $T_e\le T_0$, $T_f\le T_0$, contains 107 galaxies with a 
single star-forming region and 14 galaxies with several star-forming regions. 
We assume that $T_e$ and $T_f$ match each other in this case. 
Zone B corresponds to the case with $T_e\le T_0$, $T_f>T_0$. 
The estimations only indicate that $T_e\le T_0$, therefore the direct 
comparison is not possible. In Fig.~\ref{f5} we show the $T_f$ distribution 
for zones A and B. The left bin corresponds to the zone A with $T_f\le T_0$.
Grey and black regions correspond to the galaxies with a single star-forming 
region (subsample VI) and to the galaxies with several star-forming regions,
respectively.
 
\begin{figure}[tb]
\includegraphics[width=\columnwidth]{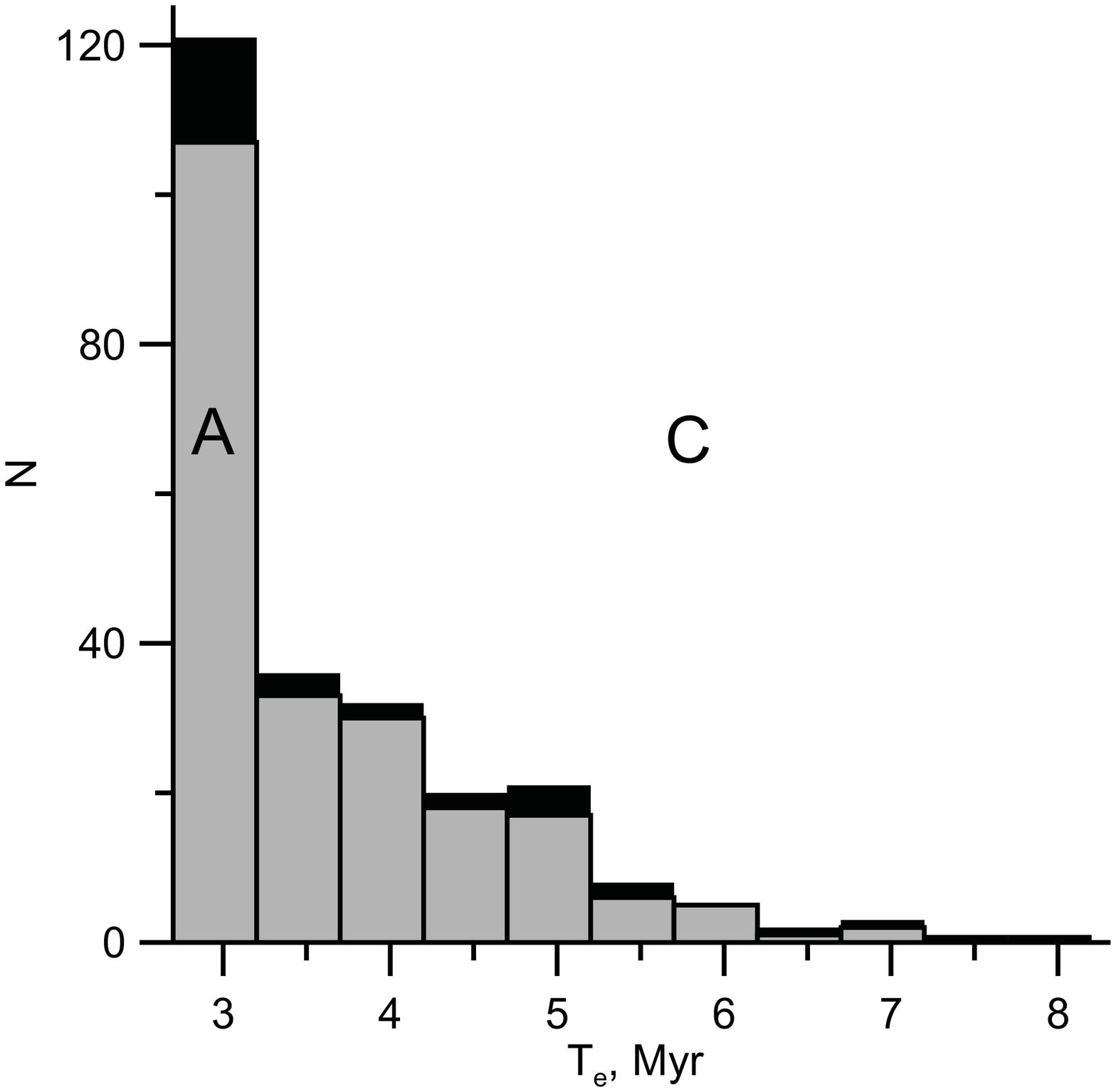}
\caption{The $T_e$ distribution in the case $T_f\le T_0$ (corresponding
to zones A and C in Fig. \ref{f4}). Grey-shaded histogram corresponds to the 
subsample VI of galaxies with a single star-forming 
region and black-shaded histogram to the galaxies with several star-forming 
regions. The whole bar (grey plus black) corresponds to the whole sample}
\label{f6}
\end{figure}

Zone C corresponds to the case with $T_f\le T_0$, $T_e>T_0$. 
In Fig. \ref{f6} we show the $T_e$ distribution for zones A and C. The left 
bin corresponds to the zone A with $T_e\le T_0$. One can see that the mean 
difference between $T_e$ and $T_f$ is greater for galaxies with several 
star-forming regions. All estimations with  $T_e\ge 7.1$ Myr refer to these
galaxies only.

Considering the alternative version of the
method, described in Section \ref{s:Alternative}, we obtain
very poor results. The values of $\widetilde{T}_e$ are
overestimated. Moreover, numerical solutions of Eqs. \ref{e6}
and \ref{e9} can be obtained only for a part of the sample. If
the data scatter were small, both methods would give similar
estimations. However, the situation changes in the case of 
significant scattering. Using the same set of the coefficients
in Eq. \ref{e4} and \ref{e8}, we obtain $\widetilde{U}=U/m^2$.
The set of parameters from Table \ref{t1} minimizes $U$, but we
reach the minimum of $\widetilde{U}$ at another values of
parameters with greater value of $\widetilde{m}_e$. The values of
SFR$_i$ fix the combination of parameters in Eq. \ref{e2}. Thus, the increase
of $m$ leads to the decrease of $\widetilde{A}\widetilde{f}(t)+\widetilde{B}\Delta$, i.e. the decrease of
$\widetilde{f}(t)$ and the increase of $t$ and
$\widetilde{T}_e=T_0+t$. However, the approximation Eq. \ref{e2}
cannot be applied to large starburst ages, when the combination
$\widetilde{A}\widetilde{f}(t)+\widetilde{B}\Delta$ changes its sign for LCGs with $\Delta>0$. As a
result, Eqs. \ref{e5} and \ref{e6} have a discontinuity of the
second kind and therefore it is not possible to find any numerical solution
for many LCGs. For this reason hereafter we will use only the method described 
in Section \ref{s:Three}.

Note that the small difference in statistical weights between Eqs. \ref{e4} 
and \ref{e8} results in an appreciable difference
in their application. Thus, hereafter we will use only the first method
based on minimization of Eq. \ref{e4}.

\section{The estimation of $m$ and $T$ values from the luminosities at three
wavelengths without invoking the galaxy oxygen abundances}\label{s:Without}

In order to calculate $m_e$ and $T_e$ one has to know the galaxy H$\alpha$, 
FUV and NUV luminosities and the oxygen abundances. However, sometimes
the accurate values of the oxygen abundances are not known. We find that  
using of the mean oxygen abundance [O]$_0$ instead of [O]$_i$ 
does not lead to the worse accuracy of the $m_e$ and $T_e$ estimation. 
An algorithm of such estimation differs from the one described in 
Section \ref{s:Three} 
by only one detail. Instead of Eq. \ref{e2} we use the version with $B=0$ and 
substitute this value as $B_i$ in Eqs. \ref{e5} and \ref{e6}. Then we obtain 
the remaining parameters, 
namely $p_i=(0.70, 0.28, 0.19)$ and $A_i=(1.98\times 10^{-7},
0.70\times 10^{-7}, 3.16\times 10^{-7})$ for the subsample VII with UV flux errors less than 
20\%. These approximations with $B_i=0$ have the mean errors $\sigma_i=(3.0, 4.2, 4.7)$ 
and are less accurate than the ones described in Section \ref{s:Parameters}. 
Nevertheless, with using these approximations we
can estimate $m$ and $T$ with errors not exceeding the ones derived with  
taking into account oxygen abundances for individual objects.

We find that the typical value $q\approx 0.28$ (Eq.~\ref{e12}) for 
the estimation without knowledge of [O] values is the same as that
with taking into account the oxygen abundance. 
The difference $T_e-T_f$ is not changed as well.  The mean-root-square values 
of these differences for the subsamples VI and VII are
0.86 Myr and 0.82 Myr, respectively.

On the other hand, we note that
discarding the galaxy metallicities from the parameter set increases the 
scatter of the approximated SFRs which are derived from H$\alpha$ line and UV 
luminosities, resulting in higher $\sigma_i$ as compared to the case considered
in Section \ref{s:Parameters}. 
Therefore, we prefer the approximation which takes into account accurate 
oxygen abundances for individual galaxies, despite the fact that it does not 
improve appreciably the determination of $m$ and $T$.

We note that the analysis in this Section is done for a sample of 
low-metallicity dwarf galaxies with [O] = 7.52 - 8.47, presently experiencing 
strong bursts of star formation. 
It may not be applicable for higher-metallicity giant galaxies and/or
galaxies with low and modest star formation. 

\section{The estimation of $m$ and $T$ values from the luminosities at two
wavelengths}\label{s:Two}

We consider now the accuracy of the $m$ and $T$ estimation if only 
H$\alpha$ and FUV luminosities are used.
From Eq. \ref{e2} we obtain two relations for SFR$_1$ and SFR$_2$, containing 
two unknown parameters $m$ and $T$. The values of $A_i, B_i$ and $p_i$ are the 
ones obtained in Section \ref{s:Parameters}. Assuming $T>T_0$, i.e. $t>0$, we 
obtain the equation
\begin{equation}\label{e12}
\begin{array}{l}
SFR_1\left( A_2\exp(-p_2t)+B_2\Delta \right)=\\
SFR_2\left( A_1\exp(-p_1t)+B_1\Delta \right).
\end{array}
\end{equation}

By solving it numerically we obtain the value of $t$. For $t>0$ we have an 
estimation of $T_e=T_0+t$ and additionally the value of $m$ from Eq. \ref{e10}.
In this case $m_1=m_2=m_e$.
The values of $m_1$ and $m_2$ are equal because of the condition in 
Eq. \ref{e12}. For $T<3.2$ Myr, corresponding to $t<0$, we adopt
$f_i=1$ and obtain two different estimations of $m_1$ and $m_2$ from 
Eq. \ref{e10}. We choose a weighted average
\begin{equation}\label{e13}
m_e=\frac{m_1\sigma_1^{-2}+m_2\sigma_2^{-2}}{\sigma_1^{-2}+\sigma_2^{-2}}
\end{equation}
as the estimated value of $m_e$.

We estimate the values of $m$ and $T$ for galaxies from their H$\alpha$ and 
FUV luminosities and oxygen abundance for the subsample VII by solving 
Eq. \ref{e12} numerically for 423 galaxies out of 425. 
The mean difference between the values of $m$ from the estimation and the 
fitting is characterized by the value $q=0.55$. Therefore,
their typical ratio is about $10^{0.55}\approx 3.55$. The mean difference 
between the values of $T$ from the estimation and from the fitting is 1.1 Myr 
and it drops to 0.9 Myr for the subsample with the errors in FUV 
fluxes lower than 10\%. As expected, the estimations based on
two wavelength bands are less accurate than the ones based on
three wavelength bands.

Estimating the values of $m_e$ and $T_e$ from the
H$\alpha$ line and NUV luminosities we obtain the mean value of $q$ equal 
to 0.5 and the mean difference of starburst ages equal to 0.83 Myr. This 
estimation is more accurate compared to the one from the H$\alpha$ line and FUV
luminosities due to a larger difference in the values of $p_i$.

Finally, estimating the values of $m_e$ and $T_e$ from their FUV and NUV 
luminosities we derive the mean value of $q$ equal to 0.57 and mean difference 
of starburst ages equal to 1.8 Myr.
However, in this case we cannot calculate the estimated values for about 10\% 
galaxies of the sample due to the lack of the data in FUV and/or NUV bands.

\section{Summary}\label{s:Summary}

We propose a new 
method for estimation of the mass $m$
and the age $T$ of the young stellar population
in Luminous Compact Galaxies (LCGs) from their
H$\alpha$ emission line, {\sl GALEX}
far-UV (FUV) and near-UV (NUV) continuun luminosities
as well as their oxygen abundances. No
time-consuming fitting of spectral energy distribution (SED) 
is needed for this purpose.

This method was validated on the sample of
about 800 LCGs constructed by \citet{I11} and used by \citet{PII}.
The typical deviations of $m$ and $T$ values derived with technique 
proposed in this paper from those obtained by \citet{I11} from the SED fitting
is 0.27 for $\log m$ and 0.87 Myr for $T$.

The proposed method
provides fairly accurate estimations for galaxies with a single
star-forming region and is slightly more accurate for galaxies
without evident asymmetry. It fails to reproduce fitted $m$ and $T$ for
galaxies with several distinguishable knots of star formation.

Adopting $k$ of $7.9 \times 10^{-42}$ for the H$\alpha$ luminosity and
of $1.4 \times 10^{-28}$ for the FUV and NUV ranges, we recommend 
the following values of the parameters to use in the determination of
$m$ and $T$: $\sigma_i=(3.0, 4.2, 4.7)$, 
$p_i=(0.72, 0.35, 0.27)$, $A_i=(2.06\times 10^{-7},
1.04\times 10^{-7}, 1.43\times 10^{-7})$ and $B_i=(-4.72\times 10^{-8},-6.21\times 10^{-8},
-9.21\times 10^{-8})$. 

We find that for galaxies with low metallicities one can use the mean 
[O] value instead of individual one without worsening the accuracy of the 
$m$ and $T$ estimation.

We also considered the estimation of $m$ and $T$ from the luminosities at two 
wavelengths.
It is less accurate, especially for estimation of $m$. The typical error of 
$\log m$ is more than $0.5$ in this case.

The proposed method can be used for preparing the samples of galaxies with the 
desired ranges of discussed parameters, preliminary selection of galaxies 
entering some samples
or for reducing of a range of parameter variations for SED fitting.

\acknowledgements

We thank the anonymous referee for valuable comments which helped
to improve the presentation of results.

{\sl GALEX} is a NASA Small Explorer launched in April 2003. We gratefully
acknowledge NASA's support for construction, operation and science analysis
for {\sl GALEX} mission, developed in cooperation with the Centre National
d'Etudes Spatiales of France and the Korean Ministry of Science and Technology.

Funding for the Sloan Digital Sky Survey (SDSS) and SDSS-II has
been provided by the Alfred P. Sloan Foundation, the Participating
Institutions, the National Science Foundation, the U.S. Department
of Energy, the National Aeronautics and Space Administration, the
Japanese Monbukagakusho, and the Max Planck Society, and the
Higher Education Funding Council for England.

\end{document}